\documentclass[twocolumn,preprintnumbers,amsmath,amssymb,superscriptaddress]{revtex4}

\usepackage{amssymb}
\usepackage{graphicx}
\usepackage{dcolumn}
\usepackage{bm}
\usepackage{amsmath}

\begin{document}


\title{Coherent superconducting quantum pump}

\author{F. Hoehne}
\email[corresponding author: ]{hoehne@wsi.tum.de}
\affiliation{Walter Schottky Institut, Technische Universit\"{a}t
M\"{u}nchen, Am Coulombwall 4, 85748 Garching, Germany}
\affiliation{RIKEN Advanced Science Institute, 34 Miyukigaoka,
Tsukuba, Ibaraki 305-8501, Japan}
\author{Yu. A. Pashkin}
\email[on leave from Lebedev Physical Institute, Moscow 119991,
Russia]{}
\affiliation{RIKEN
Advanced Science Institute, 34 Miyukigaoka, Tsukuba, Ibaraki
305-8501, Japan}
\affiliation{NEC Nano Electronics Research Laboratories,
34 Miyukigaoka, Tsukuba, Ibaraki 305-8501, Japan}
\author{O. V. Astafiev}
\affiliation{RIKEN
Advanced Science Institute, 34 Miyukigaoka, Tsukuba, Ibaraki
305-8501, Japan}
\affiliation{NEC Nano Electronics Research Laboratories, 34
Miyukigaoka, Tsukuba, Ibaraki 305-8501, Japan}
\author{M. M\"{o}tt\"{o}nen}
\affiliation{Department of Applied Physics/COMP, Aalto University,
POB 14100, 00076 Aalto, Finland} \affiliation{Low Temperature
Laboratory, Aalto University, POB 13500, 00076 Aalto, Finland}
\author{J. P. Pekola}
\affiliation{Low Temperature Laboratory, Aalto University, POB
13500, 00076 Aalto, Finland}

\author{J. S. Tsai}
\affiliation{RIKEN
Advanced Science Institute, 34 Miyukigaoka, Tsukuba, Ibaraki
305-8501, Japan}
\affiliation{NEC Nano Electronics Research Laboratories, 34
Miyukigaoka, Tsukuba, Ibaraki 305-8501, Japan}

\date{\today}

\begin{abstract}
We demonstrate non-adiabatic charge pumping utilizing a sequence of coherent oscillations between a superconducting island and two reservoirs. Our method, based on pulsed quantum state manipulations, allows to speedup charge pumping to a rate which is limited by the coupling between the island and the reservoirs given by the Josephson energy. Our experimental and theoretical studies also demonstrate that relaxation can be employed to reset the pump and avoid accumulation of errors due to non-ideal control pulses.

\end{abstract}


\maketitle


\emph{Introduction}---As electronic circuits are scaled down in size, Coulomb blockade effects~\cite{Averin86} start
to play an important role. This offers the possibility to manipulate individual charges, either single electrons or
Cooper pairs. By utilizing the charge degrees of freedom, one can not only
demonstrate control at the level of elementary charges but also apply these
techniques for practical purposes. In particular, adiabatically operated charge turnstiles~\cite{Geerligs90, Pothier92} and pumps~\cite{Keller99}
are promising candidates for redefining the unit of the ampere in quantum metrology~\cite{Keller96}. Nevertheless, the minimum current level of
100~pA, required for a so-called quantum metrological triangle experiment~\cite{Likharev85}, was yet out of reach
with these devices. High-frequency operation of charge pumps has been demonstrated in GaAs nanostructures yielding currents of almost 100~pA~\cite{Blumenthal07, Giblin10},
while experimental~\cite{Pekola08} and theoretical~\cite{Averin08} studies of a hybrid
turnstile promise a satisfactory high yield of 100~pA.

On the other hand, nonadiabatic charge state control has been demonstrated in a number of experiments on
superconducting quantum bits (qubits)~\cite{Nakamura99, Pashkin03, Yamamoto03}. The manipulation and measurement steps constitute a
cycle during which a Cooper pair is coherently transferred through a Josephson junction to an island and then
the island is reset to the initial state by incoherent charge transfer through another junction. 

Here, we apply nonadiabatic Cooper pair control for coherent charge
transfer through a Cooper pair transistor. By applying two sequential
$\pi$-pulses to the device we transfer one Cooper pair from the
source to the island and then from the island to the drain resulting
in nonvanishing average current through the system. In contrast to already
demonstrated adiabatic Cooper pair
pumping~\cite{Niskanen, Vartiainen, Mottonen06, Mottonen08}, nonadiabatic operation is 
in principle the fastest way of pumping Cooper pairs and therefore 
produces the highest pumped current. However, this method induces pumping errors which tend to accumulate from cycle
to cycle. The nonideality of the control pulses and dephasing due to
background charge fluctuations are the main sources of these errors.
Fortunately, we find that by initializing the system after each pumping
cycle, the accumulation of errors can be avoided and a pumped dc current
is observed. The pumping efficiency of the device, although less than unity, greatly exceeds the
efficiency of the recently reported superconducting quantum pump of a different type~\cite{Giazotto11}.

\begin{figure}
\includegraphics[width=8cm]{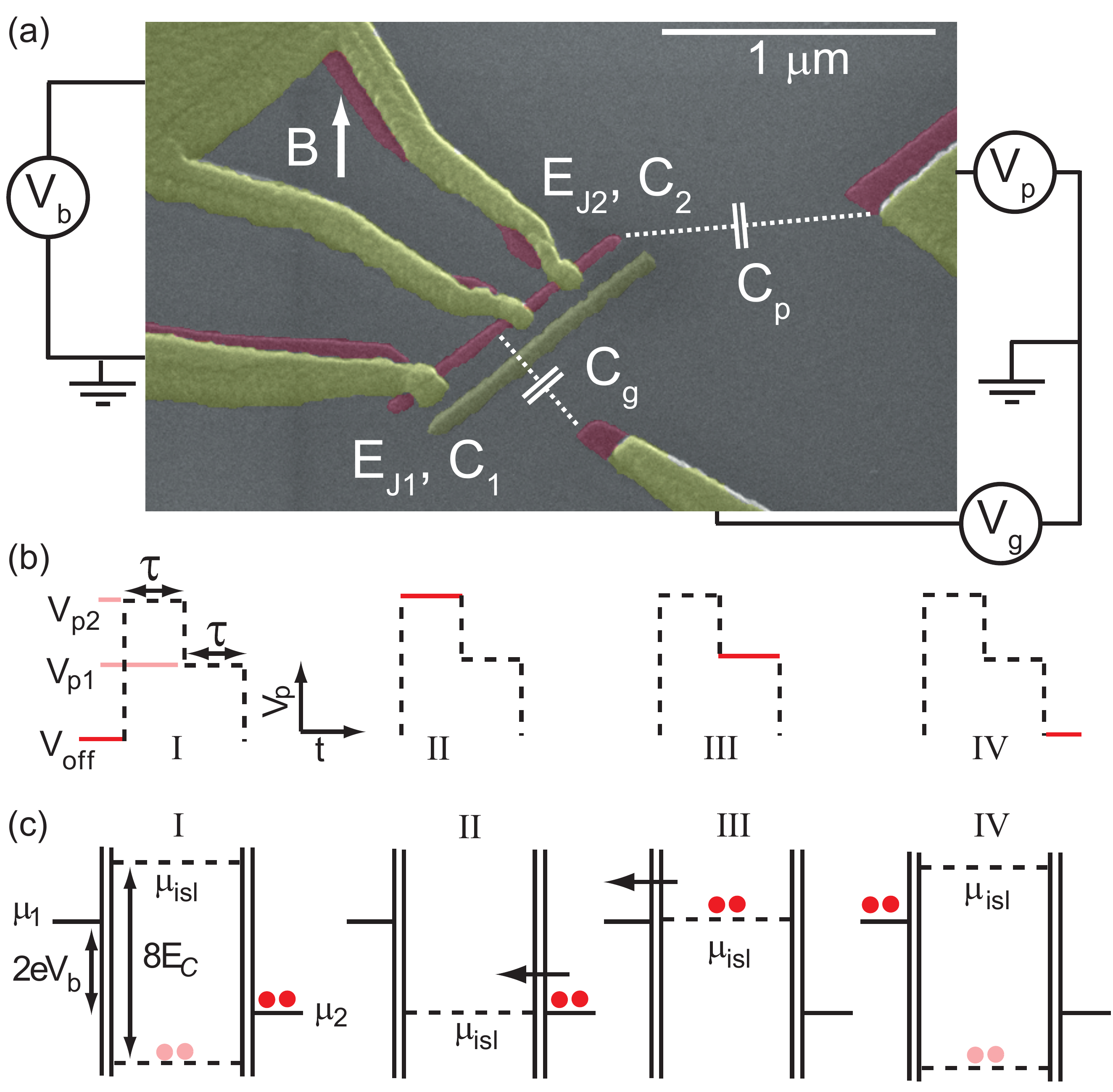}
\caption{\label{fig:Figure1}(a) Colored micrograph of the Cooper pair pump consisting of a superconducting island
shown as red bar separated by a single Josephson junction and a SQUID. The energy levels of the island
are controlled by the dc gate and the high frequency pulse gate. The basic pumping principle and the corresponding pulse sequence are depicted in
panels (b) and (c), where the solid line indicates the value
of $V_\mathrm{p}$ in each phase. With this pulse sequence, Cooper-pairs are transferred through the island against the
bias voltage. Note that a positive pulse voltage shifts the electrostatic potential of the island down.}
\end{figure}

\emph{Theoretical model}---The measured devices are based on a
Cooper pair transistor that, in addition to the dc gate, has also a
pulse gate as shown in Fig.~\ref{fig:Figure1}. The superconducting island of the
transistor (red bar) is separated on one side by a single junction with Josephson energy $E_\mathrm{J1}$
and on the other side by a superconducting quantum interference
device (SQUID). The SQUID works effectively as a single junction
with a flux-controllable Josephson energy $E_\mathrm{J2}$, which allows us to tune
$E_\mathrm{J2}$ by an external magnetic field $B$~\cite{Tinkham}. The left lead of the transistor is grounded and the right lead is biased by a voltage $V_\mathrm{b}$ (see Fig.~\ref{fig:Figure1}). A
similar device but with symmetric bias of the leads is analysed in Ref.~\cite{Maassen91}. We present the Hamiltonian of our system in the form
\begin{eqnarray}\label{eq:hami}
\hat{H}=4E_\mathrm{C}(\hat{n}-n_\mathrm{g})^2-2 eV_\mathrm{b}\hat{\bar{n}}+
\sum_{k,m}\Big(\frac{E_\mathrm{J1}}{2}|k+1, m\rangle\langle k, m| \nonumber +\\ \frac{E_\mathrm{J2}}{2} |k+1, m\rangle\langle
k, m+1|+\textrm{c.c.}\Big),
\end{eqnarray}
where the charging energy of the island $E_\mathrm{C}$ is given by the capacitance $C_1$ of lead~1 (grounded), $C_2$ of lead~2 (biased),
the gate capacitances $C_\mathrm{g}$ and $C_\mathrm{p}$, and the self capacitance of the island $C_0$ as
$E_\mathrm{C}=e^2/2(C_1+C_2+C_\mathrm{g}+C_\mathrm{p}+C_0)$. The number operators of the excess Cooper pairs on the island $\hat{n}=\sum_{k,m}k|k, m\rangle\langle k, m|$
and on lead~2 $\hat{\bar{n}}=\sum_{k,m}m|k,m\rangle\langle k,m|$ are expressed with the charge basis $|k,m\rangle$ of the number of Cooper pairs on the island ($k$), and on lead~2 ($m$).
The induced gate charge in units of $2e$ is given by $n_\mathrm{g}=(V_\mathrm{g}C_\mathrm{g}+V_\mathrm{p}C_\mathrm{p}+V_\mathrm{b}C_2)/2e$. 
The first term in the sum of the Hamiltonian~(\ref{eq:hami}) represents the Josephson coupling of the island to lead~1 and the second is coupling between the island and lead~2.

%
\emph{Pumping cycle}---The nonadiabatic pumping cycle can be realized with the composite pulse shown in
Fig.~\ref{fig:Figure1}(b) and referred to as the base sequence. Figure~\ref{fig:Figure1}(c) describes how
Cooper pairs are transferred to and from the island during the cycle: First, the electrostatic potential
of the island is brought into resonance with the second lead introducing coherent tunneling of a Cooper pair
into the island (II). Then the potential is shifted into resonance with the first lead, through which
the excess Cooper pair coherently tunnels out (III).

To describe the pumping cycle, we assume that the system with $E_\mathrm{C}\gg E_\mathrm{J1}=E_\mathrm{J2}=E_\mathrm{J}$ is initialized into state
$|00\rangle$, and that a bias voltage $0<V_\mathrm{b}\lesssim E_\mathrm{C}/e$ between the leads is applied. We nonadiabatically shift the gate charge from the point
$n_\mathrm{g}\approx 0$ (I) which is far away from charge degeneracy
to the value $n_\mathrm{g}=\frac{1}{2}+(eV_\mathrm{b})/4E_\mathrm{C}$ (II). At this
point, the states $|0,0\rangle$ and
$|1,-1\rangle$ are degenerate and the Hamiltonian~(\ref{eq:hami}) reduces to
$E_\mathrm{J}(|0,0\rangle\langle 1,-1|+|1,-1\rangle\langle 0,0|)/2$. 
By choosing the pulse
length at this level as $\tau=\pi\hbar/E_\mathrm{J}$, i.e., a $\pi$-pulse, the initial state $|0,0\rangle$ changes to
$|1,-1\rangle$ by coherent tunneling of a Cooper pair through the second junction. During the second part of the
pulse, we nonadiabatically shift the gate charge to $n_\mathrm{g}=1/2$ (III), where the effective Hamiltonian is $E_\mathrm{J2}(|1,-1\rangle\langle 0,-1|+|0,-1\rangle\langle 1,-1|)/2$, in order to induce coherent oscillations through the first junction. After the interval
$\tau=\pi\hbar/E_\mathrm{J2}$, the charge state $|1,-1\rangle$ is transferred into $|0,-1\rangle$ (IV). 
Thus, the charge transfer process induced in the whole cycle is $|0,0\rangle \rightarrow |1,-1\rangle \rightarrow |0,-1\rangle$. 
Repeating the manipulation sequence one can obtain states
$|0,m\rangle$ with any $m$. Hence, ideally one obtains an average dc current of
$I_\mathrm{p}=-eE_\mathrm{J}/\pi\hbar$. To pump forward, i.e., along the bias voltage, we can reverse the
order of the pulse heights $V_\textrm{p1}$ and $V_\textrm{p2}$, which results in transferring a Cooper pair from
lead~1 to lead~2. In our experiments, gating errors prevent us from
making many repetitions, and the true pumped current is determined by the waiting time in between the pulse
sequences as discussed in the following.

\emph{Experimental methods}---The device is fabricated by two-angle evaporation of Al with a thickness of
10~nm for the island (red patterns in Fig.~\ref{fig:Figure1}) and 40~nm for the leads (yellow patterns in Fig.~\ref{fig:Figure1})
on an oxidized silicon substrate using a standard trilayer resist
structure. The pattern is defined by electron-beam lithography in the top polymethylmetacrylate resist and
then transferred into a Ge layer by reactive ion etching. The lead and gate electrodes are connected via
filtered twisted-pair dc lines to room-temperature electronics for biasing and current amplification. The
pulse gate is connected to the central line of the prefabricated gold-patterned on-chip coplanar waveguides.
The waveguide is ribbon bonded to a coaxial line attenuated by 20~dB at 4~K. Composite pulses are generated
by superimposing two channels of a picosecond pulse pattern generator. The sample is mounted in vacuum in a
dilution refrigerator with a base temperature of about 30~mK. We extracted the following parameter
values for the sample studied in this work: $E_\mathrm{C}= 139\textrm{ $\mu$eV}$, $C_\mathrm{g}=3.3\textrm{ aF}$, and $E_\mathrm{J1}=E_\mathrm{J2}=
26~\mu$eV.

\begin{figure}
\includegraphics[width=8.3cm]{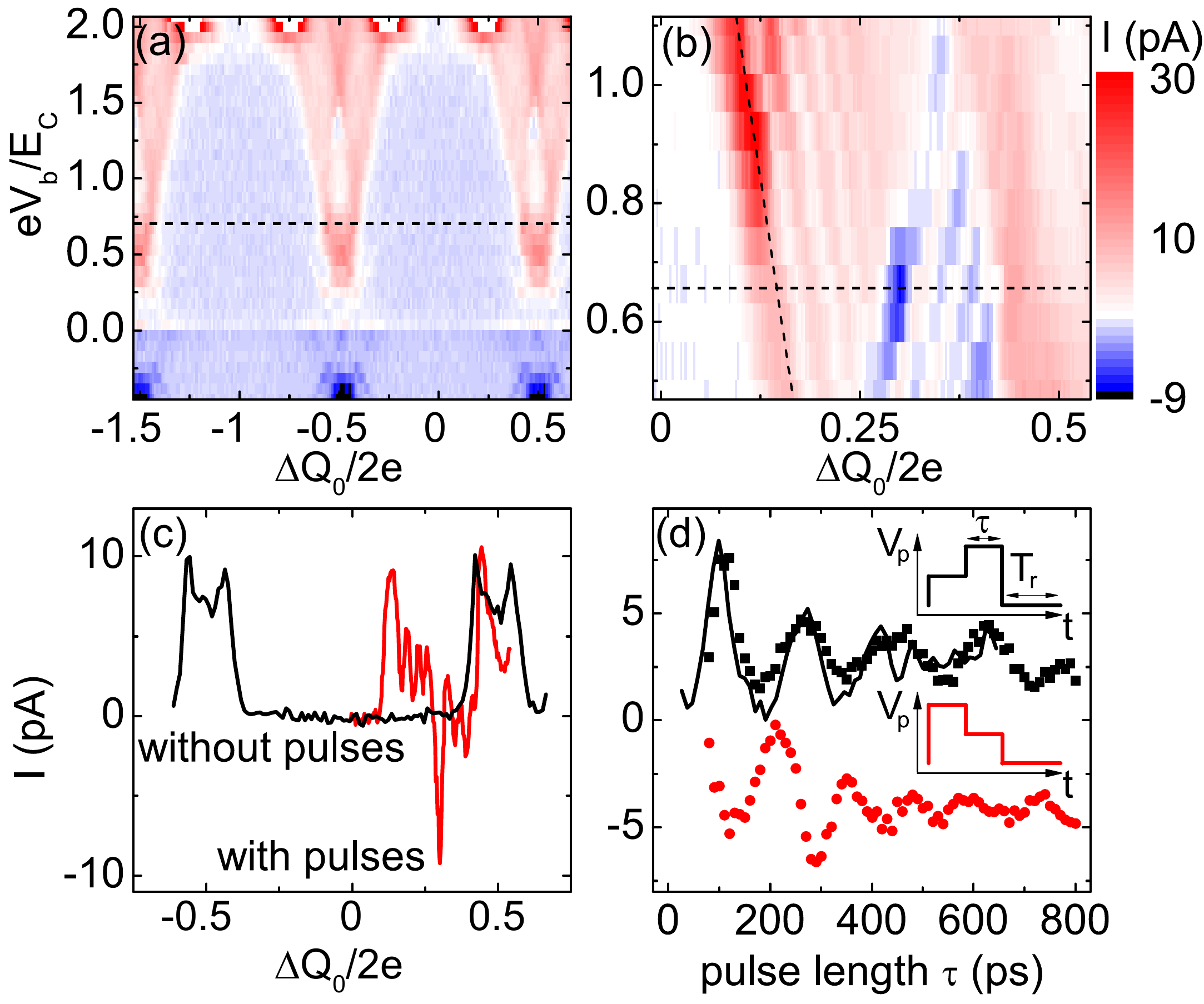}
\caption{\label{fig:Figure2} Current through the device as a function of the gate induced charge $\Delta
Q_0/2e=V_\mathrm{g}C_\mathrm{g}/2e$ and the bias voltage $V_\mathrm{b}$ (a) without and (b) with the pumping sequence applied. At $\Delta
Q_0/2e=0.32$, pumping of Cooper pairs is observed. For a direct comparison, cuts at $eV_\mathrm{b}/E_\mathrm{C}=0.66$ from
panels (a) and (b) are depicted in panel (c). (d) Pumped current as a function of the pulse length $\tau$ for
forward (black squares) and backward (red squares) pumping. The coherent oscillations have a period of 160~ps
and decay on a time scale of hundreds of picoseconds. Here, we employed $T_\mathrm{r}=8\textrm{ ns}$, $eV_\mathrm{b}/E_\mathrm{C}=0.46$, 
and $n_\textrm{p2}-n_\textrm{p1}=2eV_\mathrm{b}/8E_\mathrm{C}$. The insets show the continuously repeated
pumping sequences in each case with the number of base sequences $n=1$. The continuous lines are simulations
based on the Hamiltonian~(\ref{eq:hami}). (See text for details.)}
\end{figure}

\emph{Results}---The current through the device without applying the pumping sequence is shown in
Fig.~\ref{fig:Figure2}(a) as a function of the bias voltage $V_\mathrm{b}$ and the dc gate-induced charge $\Delta Q_0/2e=V_\mathrm{g}C_\mathrm{g}/2e$ controlled by $V_\mathrm{g}$.
Around $V_\mathrm{b}=0$, a 2$e$-periodic supercurrent is visible, confirming that our
device is not poisoned by quasiparticles. At higher bias voltages Cooper-pair tunneling resonances become
energetically allowed, accounting for some of the other features in Fig.~\ref{fig:Figure2}(a). In particular,
the V-shaped regions around the charge degeneracy points originate from resonant tunneling of one Cooper-pair
on or off the island. The strong 1e-periodic features at $eV_\mathrm{b}=2E_\mathrm{C}$ occur at the crossing of two such
Cooper-pair tunneling resonances~\cite{LepThunPRB2008, BillPRL2007}.

For Cooper pair pumping, we utilize the two-level base sequence discussed above and shown in
Fig.~\ref{fig:Figure1}(b), but in each cycle we apply $n$ subsequent base sequences followed by a waiting
period with length $T_\mathrm{r}$ at voltage $V_\mathrm{p}=V_\textrm{off}$ to allow the system to relax back to the ground
state. The current through the device with the pumping cycles applied is shown in Fig.~\ref{fig:Figure2}(b)
as a function of the bias voltage and the dc gate-induced charge $\Delta Q_0/2e$ for $n=1$,
$T_\mathrm{r}=8~\textrm{ns}$, and the pulse duration $\tau=100~\textrm{ps}$. The pulse levels at the pulse generator
are $V_\mathrm{p1}=0.8$~V and $V_\mathrm{p2}=2$~V. The corresponding dimensionless gate induced charges defined according to $n_{\mathrm{p}i} = V_{\mathrm{p}i} C_\mathrm{p}/2e$ ($i$=1,2) are $n_\mathrm{p1} = 0.11$ and $n_\mathrm{p2} = 0.28$. In addition to the Cooper-pair tunneling resonance current
($\Delta Q_0/2e=0.5$) observed also without pulses, a positive current peak ($\Delta Q_0/2e\approx0.13$)
and a negative current peak ($\Delta Q_0/2e\approx0.3$) are observed. For better visibility, a cut along the
$\Delta Q_0/2e$-axis for $eV_\mathrm{b}/E_\mathrm{C}=0.66$ is shown in Fig.~\ref{fig:Figure2}(c). We attribute
the positive current peak to a process, in which an excess Cooper pair tunnels coherently to the island from
the first lead during the pulse level $V_\textrm{p2}$ (see Fig.~\ref{fig:Figure1}) and then relaxes
incoherently to the second lead during the waiting period. According to this interpretation, the positive
current peak should appear at a gate-induced charge $\Delta Q_0/2e=1/2-n_\mathrm{p2}-e V_\mathrm{b}/8E_\mathrm{C}$. We use this relation to find the correspondence between $V_\mathrm{p}$ and $n_p$ by measuring the position $\Delta Q_0/2e$ of the positive current peak as
a function of $V_\mathrm{p2}$ while keeping the bias voltage fixed. We find that $V_\mathrm{p}=1$~V corresponds to $n_\mathrm{p}=0.14$. Using
this calibration, the expected position of the positive current peak can be calculated as shown by the dashed black
line in Fig.~\ref{fig:Figure2}(b). The good agreement with the experimental data corroborates our
interpretation of the transport process giving rise to this peak. 

Since the pumping cycle introduced in Fig.~\ref{fig:Figure1}(b) and (c) produces a negative current, we
attribute the negative current peak in Fig.~\ref{fig:Figure2}(b) to pumping. This claim is supported by the
fact, that pumping is effective only when the pulse amplitude $V_\mathrm{p2}-V_\mathrm{p1}$ corresponds to
the difference in the potentials between the leads given by $\mu_1-\mu_2=2eV_\mathrm{b}$. For
$(V_\mathrm{p2}-V_\mathrm{p1})=1.2$~V, pumping is therefore expected to be
effective at a bias voltage $eV_\mathrm{b}/E_\mathrm{C}=4(n_\mathrm{p2}-n_\mathrm{p1})=0.66$
in good agreement with the data in Fig.~\ref{fig:Figure2}(b), where the positive current peak is visible for
all bias voltages but the negative current is peaked near $eV_\mathrm{b}/E_\mathrm{C}=0.66$. In addition, the position
$\Delta Q_0/2e$ of the pumping peak should be at $\Delta
Q_0/2e=1/2-n_{\mathrm{p}1}-eV_\mathrm{b}/8E_\mathrm{C}=0.31$ close to $\Delta
Q_0/2e=0.32$ as observed in the experiment giving further support to our assignment of the negative current
peak to pumping. We have repeated the measurements shown in Fig.~\ref{fig:Figure2}(b) for pulse amplitudes
ranging from $V_\mathrm{p2}-V_\mathrm{p1}=0.5$~V to 2~V giving similar results consistent with the
interpretation described above (data not shown).

%
%

To demonstrate that the Cooper pair pumping is coherent, we measured the pumped current as a function of the
pulse length $\tau$ as shown in Fig.~\ref{fig:Figure2}(d). The bias voltage is set to
$eV_\mathrm{b}/E_\mathrm{C}=0.46$ and the corresponding pulse amplitudes are set to $V_\mathrm{p1}=0.75$~V
and $V_\mathrm{p2}=1.5$~V. We obtain a negative current with the base sequence shown in Fig.~\ref{fig:Figure1}
and a positive current with a similar sequence but with the order of the pulse levels $V_\textrm{p1}$ and
$V_\textrm{p2}$ reversed [insets in Fig.~\ref{fig:Figure2}(d)].
In both cases, oscillations of the current as a function of the
pulse length $\tau$ are observed as expected for the Hamiltonian
given in Eq.~\eqref{eq:hami}. The oscillations decay on the time
scale of hundreds of picoseconds, faster than previously
observed in charge qubits~\cite{Nakamura99}. This decay is dominated
by background charge fluctuations which change the resonance
condition for the leads, and hence imply rather fast dephasing of
the Cooper pair oscillations through the junctions, as confirmed by
our numerical simulation of the driven quantum evolution [black line in Fig.~\ref{fig:Figure2}(d)].
%
\begin{figure}
\includegraphics[width=7.3cm]{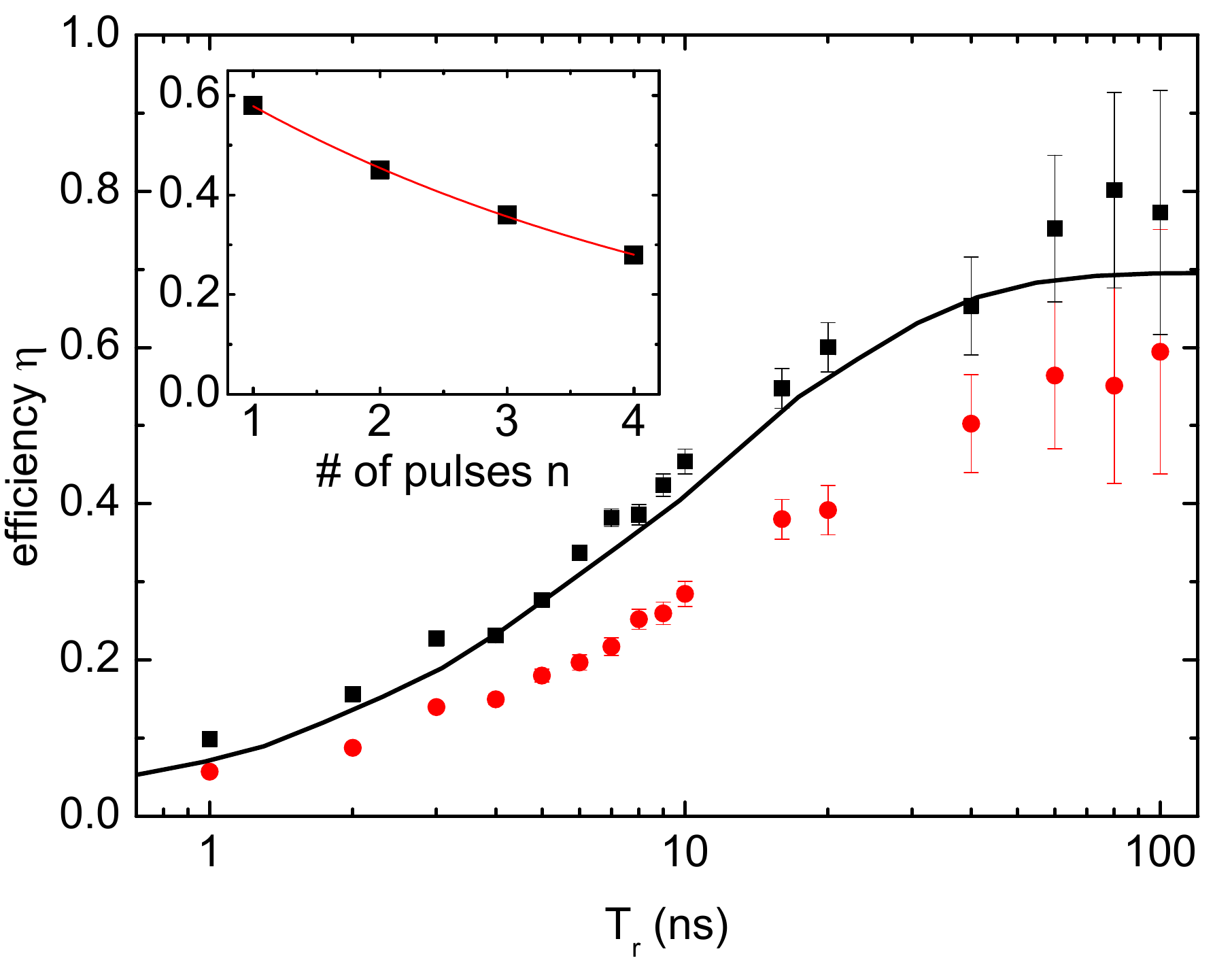}
\caption{\label{fig:Figure3} Pumping efficiency
$\eta=I/I_\textrm{max}$ as a function of the pumping period $T_\mathrm{r}$
for forward and backward pumping with the maximum efficiencies
$\eta_{\rm{max}}=0.8$ and $\eta_{\rm{max}}=0.6$, respectively. The continuous lines are simulation
including an energy relaxation rate of $\Gamma_1$~=~8~ns. The maximum efficiency depicted in the inset for backward pumping shows behavior
$\eta_{\rm{max}}(n)=A \eta_0^n$, where $\eta_0=0.74$ is the
efficiency per pulse and $A$ is the efficiency independent of the
number of pulses.}
\end{figure}
%
\newline
\indent For a Cooper pair pumping sequence, composed of $n$ base sequences, the maximal expected current is
given by $I_{\rm{max}}=2en/T_\mathrm{r}$, since $\tau\ll T_\mathrm{r}$. Thus we characterize the pumping efficiency by
$\eta=I/I_{\rm{max}}$, where $I$ is the actual pumped current. In Fig.~\ref{fig:Figure3}(a), the efficiency
for forward and backward pumping with $n=1$ is shown as a function of $T_\mathrm{r}$. The efficiency for both
directions of pumping approaches exponentially the maximal efficiency
$\eta_{\rm{max}}$ with increasing $T_\mathrm{r}$. This dependence can be phenomenologically described by
$\eta(T_\mathrm{r})=\eta_{\rm{max}}(1-e^{-T_\mathrm{r}/T})$, where $T\approx10$~ns is a characteristic time constant which is of the order of the energy relaxation time found in previous experiments for charge
qubits~\cite{Astafiev04}. The maximum efficiencies are $\eta_{\rm{max}}=0.8$ and $\eta_{\rm{max}}=0.6$ for
forward and backward pumping, respectively. Due to the accumulation of pumping errors, this efficiency decreases
for larger $n$ of base sequences in a cycle as shown in the inset of Fig.~\ref{fig:Figure3} up to $n=4$ for backward
pumping. The maximum efficiency $\eta_{\rm{max}}$ is observed to be proportional to $\eta_0^n$, where
$\eta_0=0.74$ is the efficiency per pulse.
\newline
\indent The different efficiencies for forward and backward pumping in Fig.~\ref{fig:Figure3} can be
attributed to the finite rise time of the pump pulses. In the case of backward pumping, the energy level of
the island is swept through the degeneracy point resulting in a possible tunneling process of a Cooper pair from the right lead to the island (Fig.~\ref{fig:Figure1}). Since this process transfers Cooper pairs in the direction of the applied bias voltage, the effective current for the backward pumping is decreased.
\newline
\indent 
\emph{Conclusion}---We have introduced a device for nonadiabatic Cooper pair pumping and demonstrated its
working principles both theoretically and experimentally. Due to accumulation of pumping errors, the average
pumped current was found to be determined by the internal relaxation rate of the device rather than the
Josephson energy. Although more sophisticated, error correcting, pumping sequences may improve the operation,
it remains to be shown whether nonadiabatic operation provides advantage over adiabatic Cooper pair
pumping~\cite{Niskanen,Vartiainen}. In future, it would be interesting to study the possible relation between
the geometric phases and the nonadiabatically pumped charge as has been already demonstrated in the adiabatic
case~\cite{Mottonen06, Mottonen08}.

We thank J. Kokkala for discussions.
This work was supported by MEXT kakenhi "Quantum Cybernetics," the JSPS through its FIRST Program,
the European Community's Seventh Framework Programme under Grant Agreement No. 238345 (GEOMDISS),
the Academy of Finland and Emil Aaltonen Foundation.

\clearpage

\begin{thebibliography}{99}

\bibitem{Averin86} D.\ V.\ Averin and K.\ K.\ Likharev, J.\ Low Temp.\ Phys.\ \textbf{62}, 345 (1986).

\bibitem{Geerligs90} L.\ J.\ Geerligs, V.\ F.\ Anderegg, P.\ A.\ M.\ Holweg, J.\ E.\ Mooij,
H.\ Pothier, D.\ Esteve, C.\ Urbina, and M.\ H.\ Devoret, Phys.\
Rev.\ Lett.\ \textbf{64}, 2691 (1990).

\bibitem{Pothier92}  H.\ Pothier, P.\ Lafarge, C.\ Urbina, D.\ Esteve, and M.\ H.\ Devoret,
Europhys.\ Lett.\ \textbf{17}, 249 (1992).

\bibitem{Keller99} M.\ W.\ Keller, A.\ L.\ Eichenberger, J.\ M.\ Martinis, and N.\ M.\ Zimmerman,
Science \textbf{285}, 1706 (1999).

\bibitem {Keller96} M.\ W.\ Keller, J.\ M.\ Martinis, N.\ M.\ Zimmerman, and A.\ H.\ Steinbach,
Appl.\ Phys.\ Lett.\ \textbf{69}, 1804 (1996).

\bibitem{Likharev85} K.\ K.\ Likharev and A.\ B.\ Zorin, J.\ Low Temp.\
Phys.\ {\bf 59}, 347 (1985).

\bibitem{Blumenthal07} M.\ D.\ Blumenthal, B.\ Kaestner, L.\ Li, S.\ Giblin, T.\ J.\ B.\ M.\ Janssen, M.\ Pepper, D.\ Anderson, G.\ Jones, and D.\ A.\ Ritchie, Nature\ Phys.\ {\bf 3}, 343 (2007)

\bibitem{Giblin10} S.\ P.\  Giblin, S.\ J.\ Wright, J.\ D.\ Fletcher, M.\ Kataoka, M.\ Pepper, T.\ J.\ B.\ M.\ Janssen, D.\ A.\ Ritchie, C.\ A.\ Nicoll, D.\ Anderson, and G.\ A.\ C.\ Jones, New\ J.\ Phys.\ {\bf 12}, 073013 (2010)

\bibitem{Pekola08} J.\ P.\ Pekola, J.\ J.\ Vartiainen, M.\ M\"{o}tt\"{o}nen, O.-P.\ Saira, M.\ Meschke, and
D.\ V.\ Averin, Nature Phys.\ \textbf{4}, 120 (2008).

\bibitem{Averin08} D.\ V.\ Averin and J.\ P.\ Pekola, Phys.\ Rev.\ Lett.\ \textbf{101}, 066801 (2008).

\bibitem{Nakamura99} Y.\ Nakamura, Yu.\ A.\ Pashkin, and J.\ S.\ Tsai, Nature \textbf{398}, 786 (1999).


\bibitem{Pashkin03} Yu.\ A.\ Pashkin, T.\ Yamamoto, O.\ Astafiev, Y.\ Nakamura, and J.\ S.\ Tsai, Nature \textbf{421}, 823 (2003).

\bibitem{Yamamoto03} T.\ Yamamoto, Yu.\ A.\ Pashkin, O.\ Astafiev, Y.\ Nakamura, and J.\ S.\ Tsai, Nature \textbf{425}, 941 (2003).

\bibitem{Niskanen} A.\ O. Niskanen, J.\ M.\ Kivioja, H.\ Sepp\"a, and J.\ P.\
Pekola, Phys.\ Rev.\ B \textbf{71}, 012513 (2005).

\bibitem{Vartiainen} J.\ J.\ Vartiainen, M.\ M\"ott\"onen, J.\ P.\
Pekola, and A.\ Kemppinen, Appl.\ Phys.\ Lett.\ \textbf{90}, 082102
(2007).

\bibitem{Mottonen08} M.\ M\"ott\"onen, J.\ J. Vartiainen, and J.\ P.\
Pekola, Phys.\ Rev.\ Lett.\ \textbf{100}, 177201 (2008);

\bibitem{Mottonen06} M.\ M\"ott\"onen, J.\ P.\ Pekola, J.\ J.\ Vartiainen, V.\ Brosco, and
F.\ W.\ J.\ Hekking, Phys.\ Rev.\ B \textbf{73}, 214523 (2006).

\bibitem{Giazotto11} F.\ G.\ Giazotto, P.\ Spathis, S.\ Roddaro, S.\ Biswas, F.\ Taddei, M.\ Governale, and L.\ Sorba, Nat.\ Phys.\ (2011), online only.

\bibitem {Tinkham} M. Tinkham, \textit{Introduction to {S}uperconductivity} (Dover, New York, 2004).

\bibitem{Maassen91} A.\ Maassen van den Brink, A.\ A.\ Odintsov, P.\ A.\
Bobbert, and G.\ Sch{\"o}n, Z.\ Phys.\ B: Condens. Matter
\textbf{85}, 459 (1991).

\bibitem {LepThunPRB2008} J.\ Lepp\"akangas and E.\ Thuneberg, Phys.\ Rev.\ B \textbf{78}, 144518 (2008).

\bibitem {BillPRL2007} P.-M.\ Billangeon, F.\ Pierre, H.\ Bouchiat, and R.\ Deblock,  Phys.\ Rev.\ Lett. \textbf{98}, 216802 (2007).

\bibitem{Astafiev04} O.\ Astafiev, Yu.\ A.\ Pashkin, Y.\ Nakamura, T.\ Yamamoto, and J.\ S.\ Tsai, Phys.\ Rev.\ Lett.\ \textbf{93}, 267007 (2004).



\end{thebibliography}
\end{document}